\documentstyle[12pt]{article}
\normalsize
\def\hc1{H_{c_1}}

\def\be{\begin{equation}} 
\def\ee{\end{equation}} 
\def\bea{\begin{eqnarray}} 
\def\eea{\end{eqnarray}} 
\def\nn{\nonumber}

\newcounter{sxn}

\newcounter{axn}

\def\br{\begin{thebibliography}{abc}}
\def\er{\end{thebibliography}}

\begin{document}
\bibliographystyle{unsrt}
\footskip 1.0cm
\thispagestyle{empty}
\begin{flushright}
NSF-ITP-94-21\\
March 1994\\
\end{flushright}
\vspace*{10mm}
\centerline {\Large TOWARDS OBSERVING THE INTERCOMMUTATION OF}
\vspace*{3mm}
\centerline {\Large FLUX TUBES IN SUPERCONDUCTORS} 
\vspace*{15mm}
\centerline {\large Ajit M. Srivastava}
\vspace*{5mm}
\centerline {\it Institute for Theoretical Physics, University of California}
\centerline {\it Santa Barbara, California 93106, USA}
\vspace*{15mm}
 
\baselineskip=18pt

\centerline {\bf ABSTRACT}
\vspace*{8mm}
 
 We propose a simple experiment to investigate the intercommutation
of flux tubes in type II superconductors. Using this method
the intercommutation of strings can be observed directly
and the dependence of intercommutation on the angle of crossing
of strings can also be analyzed.
 
\newpage
 

 Interaction of topological defects has been a subject of great interest.
Interaction of flux tubes in type II superconductors is not only
of theoretical interest but also has technological implications.
In many particle theory models, strings in the early Universe play
crucial role in generating density fluctuations responsible for
the structure formation. In such models interaction properties of
strings are of utmost importance. An essential property of cosmic
strings is the intercommutativity; that is when two strings cross each 
other, they exchange partners at the crossing point as shown in Fig.1. 
Due to this property, large loops of cosmic strings chop themselves
into smaller ones which rapidly decay through oscillations and
emission of radiation. If this was not so then strings would have 
dominated the energy density of the Universe very early and would be
observationally ruled out.

 There have been many theoretical studies of string intercommutativity
in relativistic field theory models. Shellard investigated string
interactions numerically and calculated the probability of intercommutation
as a function of the crossing angle and the
relative velocity of strings for global U(1) strings \cite{shlrd}. 
He showed that strings almost always intercommute even for 
velocities very close to the speed of light. For gauge strings 
intercommutativity  was demonstrated in numerical simulations by Moriarty, 
Myers and Rebbi \cite{rebbi}.
Some qualitative understanding of the  intercommutativity of global strings 
is provided in \cite{carl}. Intercommutativity for global strings has been 
seen to occur in liquid crystal experiments where one clearly sees
strings chopping in smaller loops which shrink down rapidly 
\cite{nlc}. Fluid vortices also show intercommutativity \cite{he4}.
There is some indirect evidence for intercommutation
of flux tubes in type II superconductors \cite{crs}. [Assuming that
flux tubes intercommute, it has been suggested in \cite{rsv} that the
density of strings produced in a superconducting transition can be
experimentally investigated.]
However, a systematic experimental study of the
intercommutation of strings has not been possible, primarily
due to the difficulties in controlling the crossing of the strings.

  In this letter we propose a simple experiment where the intercommutativity
of flux tubes can be directly investigated. The essential idea of
the experiment is that flux tubes in type II superconductors, which arise
due to an external magnet, tend to follow the motion of the magnet if
that magnet is gradually moved along the surface of the superconductor
\cite{move}. This provides us with a way of effectively $holding$ the ends 
of the flux tubes. All that is needed after this is to take a crossed
configuration of flux tubes and move magnets to force one bunch
of flux tubes to go through the other bunch. Strings will then either
intercommute or simply cross each other.

 Let us describe the experiment in more detail now. Fig.2a shows
the rectangular slab of superconducting material. As the mobility of
strings is crucial in this case, high T$_c$ materials are not suitable
for this experiment due to vortex pinning. Even for conventional low
T$_c$ superconductors the sample should be very clean so vortex pinning 
is as little as possible. Crossed configuration of flux tubes is produced
by two pairs of magnets. The first pair of magnets, M$_1$ and M$_2$, 
is aligned along the x axis so the
flux tubes formed due to the field of these magnets are along the x axis.
The second pair of magnets M$_3$ and M$_4$ is aligned along the y axis
leading to strings along the y axis. N and S denote the north pole and 
the  south pole respectively. These magnets need to have very small
cross-sections so that each  bundle of flux tubes is reasonably confined
initially. The initial vertical separation of the two pairs of magnets 
along the z axis should be large compared to the spread of either bundle
of flux tubes in z direction. Magnets are to be suspended so that their 
deflections can be monitored.

 Now we fix the bottom pair of the magnets (M$_3$, M$_4$) for 
simplicity and slowly move the top pair (M$_1$, M$_2$) in
the negative z direction. Let us call the distance between the x-y
planes going through the centers of the two magnet pairs as $h$.
The forces on magnets have two components. One is $F_{ext}$, due to the 
magnetic field in the region outside of the superconductor and the second 
is $F_{str}$, due to trapped flux tubes \cite{force}. As the vertical 
separation $h$ between the two pairs decreases, the two bundles of flux tubes
will also approach each other. This movement of flux tubes due to the
movement of the associated magnets is a reasonably natural thing to
expect and there is also experimental evidence to support it \cite{move}.
[When magnet is moved considerably then associated string has to eventually
move any way.]

 When $h$ is decreased then, as long as the two bundles of strings are not 
touching each other, the
only change in the force on the magnets is due to changes in $F_{ext}$, 
which is force due to the external field.  This is because force between
the two bundles is exponentially suppressed and as long as the
two bundles of strings only get displaced and do not touch each other,
$F_{str}$ should remain practically  unchanged. As $h$ decreases, the 
magnitude of $F_{ext}$ will increase continuously as the two pairs 
of magnets  are coming closer to each other and will reach maximum value
for $h$ equal to zero, when all magnets are in the same plane. As $h$ 
is decreased further to negative values (i.e. the pair M$_1$, M$_2$ 
becomes lower than the pair M$_3$, M$_4$), the magnitude of
$F_{ext}$ will start decreasing.
This is not so with $F_{str}$ however. $F_{str}$ should remain
unchanged initially as $h$ is decreased and should remain so 
until the two bundles of strings are about to touch. If the string 
bundles have roughly the same cross-section as that
of the magnets then $F_{str}$ will remain unchanged until $h$ decreases
to a value roughly equal to the diameter $d$ of the cross-section of a magnet.
[Assuming that the magnets are moved slowly enough so that the strings
follow the magnets. If the strings lag behind the magnets then $F_{str}$
may remain unchanged for even smaller values of $h$.]

  As $h$ is decreased further, the two bundles will start strongly
overlapping. Initially the two bundles may just repel each other for
a little while meaning that $F_{str}$ will not change much. 
However, eventually some strings will break at the crossing point and
will exchange partners, that is they will intercommute. These 
intercommuted strings will quickly shrink towards the edges of the sample
and will lead to a drastic re-organization of flux tube distribution inside
the sample. Fig.2b shows the situation when many strings have intercommuted 
and these strings are shrinking towards the edges of the sample.
Some strings may simply cross each other without intercommuting as shown
in Fig.2b. Thus there will be a critical  value $h_0$ of $h$ at which the 
strings will start intercommuting. This should lead to a sudden change in
the direction of $F_{str}$ as intercommuted strings shrink towards the edges.
Since the net deflection of the magnets is a combined effect of $F_{ext}$
and $F_{str}$, a sudden change in $F_{str}$ should be observed as
a sudden change in the positions of the magnets. 

 In fact, if the experiment is carried out extremely carefully then
it may be possible to record the breaking of individual string (or at
least very few strings at a time). For this one will need to consider
very tiny cross-sections of the magnets and the magnetic field only
slightly above the lower critical field $\hc1$ so that the strings
are very dilute in the bundle. Here $\hc1$ is given by \cite{park},

\be
 \hc1 = {\phi_0 \over 4 \pi \lambda^2} ln({\lambda \over \xi})
\ee

\noindent where $\phi_0 = {hc \over 2e}$ is the flux quantum,
$\lambda$ is the penetration depth and $\xi$ is the coherence length.
From the relation between the induction $B$ and the magnetic field
$H$ \cite{park} we can choose large enough value of $H$ ($\simeq 2 \hc1$)
so that $B$ is of the order of $\hc1$. Then the density of strings 
per unit surface area $n_{ext}$ due to the external magnets 
will be given by

\bea
n_{ext} & = & {\hc1 \over \phi_0} \nn \\
   & = & {1 \over 4 \pi \lambda^2} ln ({\lambda \over \xi})
\eea

 From this equation it is clear that by applying suitably weak 
external field, a dilute bundle of strings can be
obtained where strings are well separated compared to the penetration 
depth. Fig. 3 shows a hypothetical plot of
the deflection of magnets vs. $h$. This plot is intended to only show
the qualitative aspects. As the variation of $F_{ext}$ with $h$  can 
be calculated knowing the positions of the magnets, the deflections
$\theta_{ext}$ of the magnets due to only $F_{ext}$ 
can be subtracted out in plotting $\theta$ vs. $h$. 
Fig.3 is supposed to show such a normalized plot
showing the deflection $\theta$ vs. $h$ entirely due to $F_{str}$.
One way to get  $\theta_{ext}$ due to $F_{ext}$ alone is by measuring
the deflection vs. $h$ initially until the strings are about to touch.
This gives the effect of $F_{ext}$ alone, and hence $\theta_{ext}$,
for $h > h_0$ as $F_{str}$ is  expected to remain unchanged for these 
values of $h$. Due to the symmetry of the setup, $\theta_{ext}$ will be
symmetric about $h = 0$. One may then get a rough idea of $\theta_{ext}$  
for $0 < h < h_0$ by using interpolation.  Another possibility is to
measure $\theta_{ext}$ for all values of $h$ using a slab  of a 
different superconducting material, but same geometry, which has a larger
value of $\hc1$  so that no flux tubes are formed inside the 
superconductor for the magnets used.

 As we mentioned earlier, until the two bundles of strings start
touching, $F_{str}$ (and hence $\theta$) almost remains constant at 
some intial fixed  value,  as shown in Fig.3 for
$h > d$ region of the plot. As $h$ is decreased further, strings
will start overlapping. Sudden increase in $\theta$ for $h < h_0$ is
supposed to represent breaking of one (or a small number of) strings.
As $h$ decreases further, more and more strings break leading to
further sudden changes in $\theta$. [Step in $\theta$ at a given
$h$ will be governed by the number of strings which intercommute at that
value of $h$.] It is important to realize
that the jump in $\theta$ can in principle be very sharp. This is
because $\theta$ will not change much if strings simply remain entangled.
Breaking of string (and consequent intercommutation) is a discontinuous
process and should lead to a discontinuous jump in $\theta$. If 
several strings intercommute at a time then departure from
a sharp jump can arise due to the uncertainty in the number of
strings breaking. Also, as the strings start breaking, they
may affect the actual value of $F_{ext}$ leading to an error in
properly subtracting the effects of $f_{ext}$ in plotting $\theta$
vs. $h$. This also can smoothen the jump in $\theta$. Here we 
would like to mention that if deflection due to $F_{ext}$ is not subtracted
out in plotting $\theta$ vs. $h$, then the overall slope of the curve
will be modified. However the steps in $\theta$  should still be present
even if total  deflection is plotted vs. $h$.

 As $h$ is decreased further, eventually the two bundles will cross each
other completely. A fraction of the strings will intercommute in this
process leading to the maximum possible deflection $\theta_m$. This value
$\theta_m$ gives us a measure of the final fraction of strings which
have intercommuted and hence gives the probability of intercommutation. 
It is then a rather simple task to study the
probability of intercommutation vs. the angle $\phi$ of crossing of
strings. All that one needs to do is to change the direction of the alignment
of the two magnets M$_1$ and M$_2$ so that the strings due to them
make an angle $\phi$ with the strings formed by the pair M$_3$ and M$_4$.
The whole experiment can be repeated then again leading to a plot of
the deflection $\theta$ vs. $h$ (by properly subtracting the deflection
due to $F_{ext}$). The value of the maximum deflection $\theta_m$
can then be noted down as a function of varying crossing angle $\phi$.
This gives us a direct measure of the probability of intercommutation as a
function of the crossing angle $\phi$. It should be clear that the  value of
$\phi$ for which M$_1$ and M$_3$ are on the same side of the superconducting
slab (and M$_2$ and M$_4$ on the opposite side), the change in $F_{str}$
will be larger (as compared to the case shown in Fig.2) when
the strings intercommute. This is because in that case
the intercommuted strings will be able to completely shrink and get out of
the sample. From this point of view, one may take a different geometry
of the sample (such as hexagonal) so that intercommuted strings
completely shrink away for other values of $\phi$ as well.

 It is important to note that the above experiment should be carried out by 
moving magnets very slowly to guarantee that the strings are pulled along
with the magnets. It is also of great interest to study the
probability of intercommutation as a function of the relative
velocity of the two strings. One can try to achieve this by moving
down the magnet pair at different velocities keeping the crossing angle
at some fixed value. This in principle can give the probability of
intercommutation as a function of the velocity as well as the
crossing angle. However, the velocity of strings through the
superconductor is governed by many factors such as presence of
impurities and dissipation. It will be hard in general to relate
the velocity of magnet to the velocity of the flux tube in a direct manner.
Still, the string velocity should certainly be an increasing function
of the velocity of magnets. Therefore, qualitative features of the
variation in the intercommutation probability with the relative
velocity of the strings can be studied by following the procedure
outlined in the above. 

 We conclude by pointing out the two crucial aspects of the experiment
described in this letter. One is the ability to move flux tubes by
the motion of the external magnet. This gives us a way of manipulating
flux tubes almost as real strings. Secondly, the intercommutation
of strings leads to a discontinuous rearrangement in the distribution
of strings which can then be detected as a discontinuous change in the
force on magnets. It may also be possible to use real time
observation techniques employing Lorentz microscopy \cite{elec} to 
detect the rearrangement of strings after intercommutation.

\vskip .3in
\centerline {\bf ACKNOWLEDGEMENTS}
\vskip .1in

 I am very grateful to Mathew Fisher, Jim Langer, Fong Liu, Mike Stone  
and Shikha Varma for very useful Discussions. This work was supported 
by the U. S. National Science Foundation under Grant No. PHY89-04035.



\vskip 3in
\centerline {\bf FIGURE CAPTIONS}
\vskip .1in

 (1) Intercommutation of strings.

 (2) (a) shows a rectangular slab of superconductor and the
initial arrangement of the two pairs of magnets. The two bundles
of strings are well separated initially along the z axis. 
(b) String distribution after the two bundles of strings have 
crossed leading to intercommutation of many strings.
Ratio of strings which intercommuted to the ones that simply crossed
gives a direct measure of the probability of intercommutation.

 (3) This plot shows the qualitative aspects of the variation of
the deflection $\theta$ vs. the vertical separation $h$
between the two magnet pairs. In this plot, it is assumed that the 
deflection due to the force on magnets by external magnetic field 
has been  subtracted out.
  
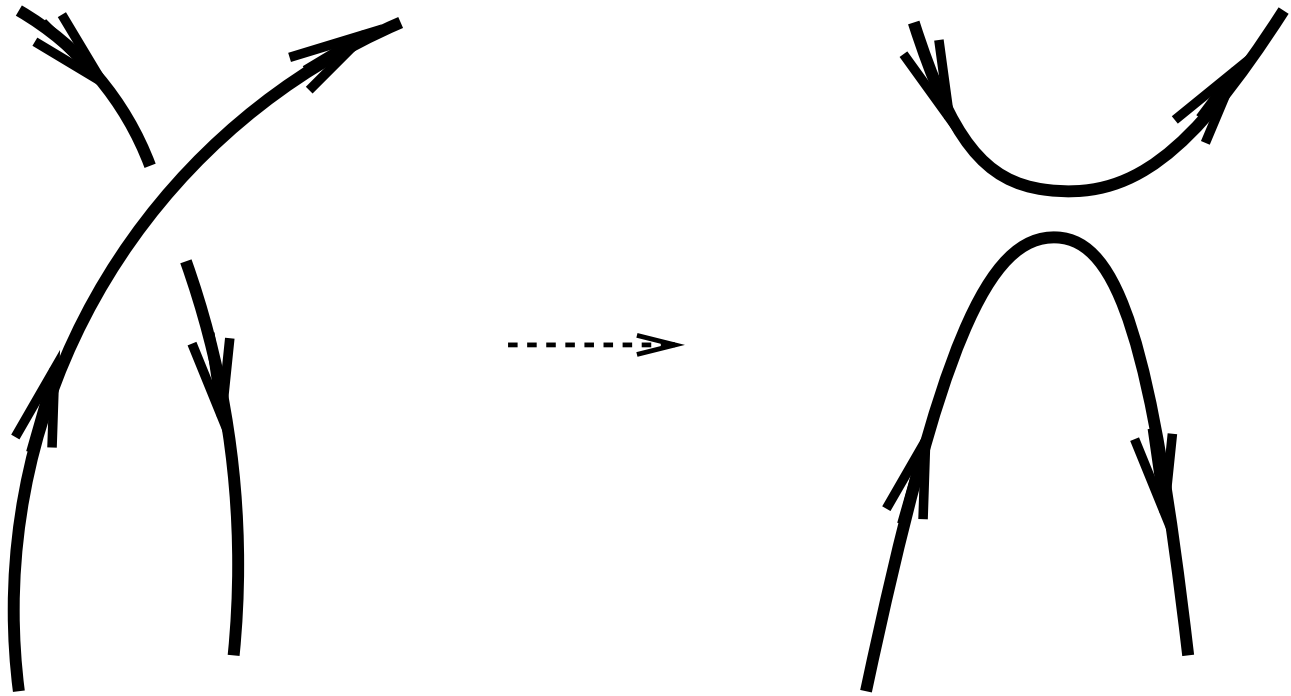
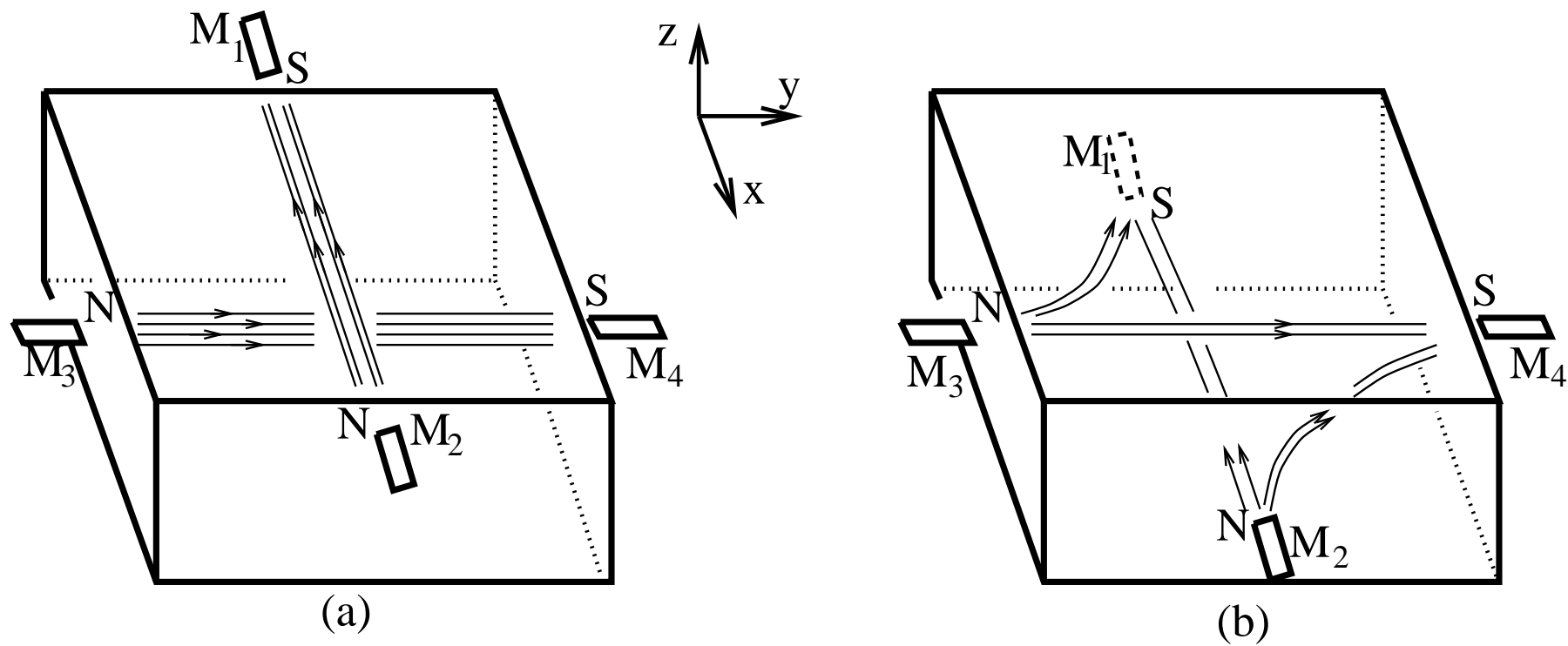
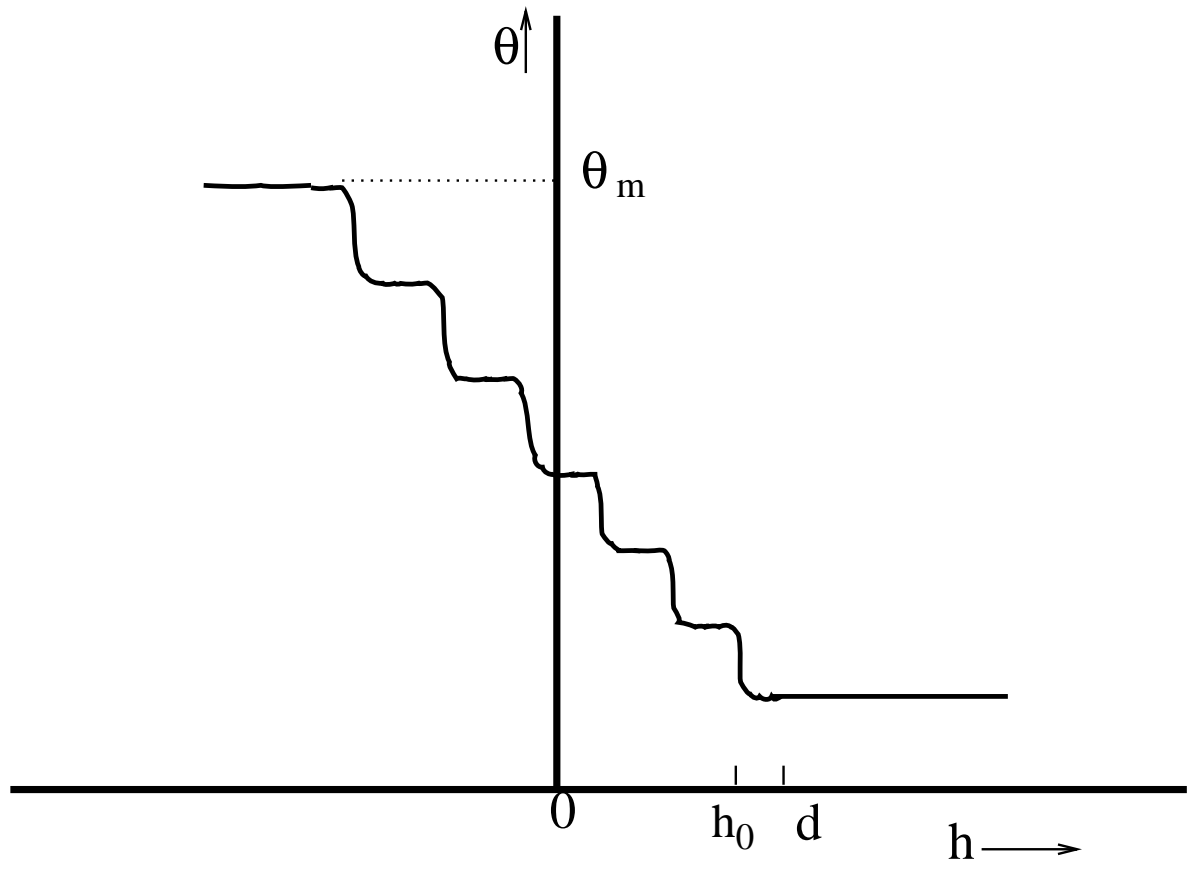
\end{document}